\documentclass[reprint,aps,prl,twocolumn,superscriptaddress,floatfix,longbibliography]{revtex4-1}

\usepackage[dvips]{graphicx}
\usepackage{latexsym}
\usepackage{amsmath}
\usepackage{amsthm}
\usepackage{amsfonts}
\usepackage{amssymb}
\usepackage{mathptmx}
\usepackage{subfigure}
\usepackage{mathrsfs}

\usepackage{color}
\definecolor{darkblue}{rgb}{0,0.02,0.45}
\usepackage[colorlinks=true,citecolor=darkblue]{hyperref}

\usepackage[all]{hypcap}
\usepackage{gensymb}

\begin{document}

\title{Gapless spin-liquid state in the structurally disorder-free triangular antiferromagnet NaYbO$_2$}

\author{Lei Ding}
\email{lei.ding.ld@outlook.com; lei.ding@stfc.ac.uk}
\author{Pascal Manuel}
\affiliation{ISIS Facility, Rutherford Appleton Laboratory, Harwell Oxford, Didcot OX11 0QX, United Kingdom}

\author{Sebastian Bachus}
\author{Franziska Gru{\ss}ler}
\author{Philipp Gegenwart}
\affiliation{Experimental Physics VI, Center for Electronic Correlations and Magnetism, University of Augsburg, 86135 Augsburg, Germany}

\author{John Singleton}
\affiliation{National High Magnetic Field Laboratory, Los Alamos National Laboratory, Los Alamos, NM 87545, United States}

\author{Roger D. Johnson}
\affiliation{Clarendon Laboratory, Department of Physics, University of Oxford, Oxford OX1 3PU, United Kingdom}

\author{Helen C. Walker}
\affiliation{ISIS Facility, Rutherford Appleton Laboratory, Harwell Oxford, Didcot OX11 0QX, United Kingdom}
\author{Devashibhai T. Adroja}
\affiliation{ISIS Facility, Rutherford Appleton Laboratory, Harwell Oxford, Didcot OX11 0QX, United Kingdom}
\affiliation{Highly Correlated Matter Research Group, Physics Department, University of Johannesburg, Auckland Park 2006, South Africa}
\author{Adrian D. Hillier}
\affiliation{ISIS Facility, Rutherford Appleton Laboratory, Harwell Oxford, Didcot OX11 0QX, United Kingdom}

\author{Alexander~A.~Tsirlin}
\affiliation{Experimental Physics VI, Center for Electronic Correlations and Magnetism, University of Augsburg, 86135 Augsburg, Germany}

\date{\today} 

\begin{abstract}
We present the structural characterization and low-temperature magnetism of the triangular-lattice delafossite NaYbO$_2$. Synchrotron x-ray diffraction and neutron scattering exclude both structural disorder and crystal-electric-field randomness, whereas heat-capacity measurements and muon spectroscopy reveal the absence of magnetic order and persistent spin dynamics down to at least 70\,mK. Continuous magnetic excitations with the low-energy spectral weight accumulating at the $K$-point of the Brillouin zone indicate the formation of a novel spin-liquid phase in a triangular antiferromagnet. This phase is gapless and shows a non-trivial evolution of the low-temperature specific heat. Our work demonstrates that NaYbO$_2$ practically gives the most direct experimental access to the spin-liquid physics of triangular antiferromagnets.
\end{abstract}

\maketitle

\textit{Introduction.} Quantum spin liquids (QSLs) in frustrated magnets have attracted a lot of attention because of the occurrence of unconventional ground states, where highly entangled spins and their strong fluctuations are observed in the absence of long-range order down to zero temperature. The ensuing excitations are interesting in their own right as they are distinct from magnons in systems with conventional long-range magnetic order~\cite{Balents2010,Savary2017,Knolle2019}.  Historically, the QSL was first exemplified by the nearest-neighbor resonating-valence-bond (RVB) state on the triangular lattice~\cite{Anderson1973}, which drew a great deal of interest in triangular antiferromagnets ever since, although the majority of real-world triangular materials order magnetically at low temperatures~\cite{Starykh2015}. 

Recently, a promising QSL candidate YbMgGaO$_4$ with the effective spin-1/2 Yb$^{3+}$ ion on triangular lattice has been proposed by Li~\textit{et al.}~\cite{Li2015SR,Li2015PRL}. They observed persistent spin dynamics down to at least 48\,mK~\cite{Li2016PRL} and the $T^{0.7}$ power-law behavior of the specific heat~\cite{Li2015SR}, indicative of the gapless U(1) QSL characterized by a Fermi surface of fractionalized (spinon) excitations. 
Although continuous spinon-like excitations were indeed observed experimentally~\cite{Shen2016}, their assignment to spinons~\cite{Shen2018} is far from unambiguous, and alternative phenomenological explanations within the valence-bond framework were proposed as well~\cite{Li2017NC,Kimchi2018}. 
Moreover, absent magnetic contribution to the thermal conductivity~\cite{Xu2016}, considerable broadening of spin-wave excitations in the fully polarized state~\cite{Paddison2017}, and acute broadening of the crystal-electric-field (CEF) excitations of Yb$^{3+}$~\cite{Li2017PRL} reveal a significant complexity of this material. The problem appears to be related to the statistical distribution of Mg$^{2+}$ and Ga$^{3+}$ that randomizes the local environment of Yb$^{3+}$~\cite{Li2017PRL} and may lead to peculiar effects like spin-liquid mimicry~\cite{Zhu2017,Parker2018}, although the exact influence of the structural randomness (disorder) on the magnetic parameters remains debated~\cite{Zhang2018}. Whether or not these structural effects are integral to the spin-liquid formation, the complexity of YbMgGaO$_4$ hinders its use as a reference model material for the QSL state in triangular antiferromagnets.

On the theory side, significant efforts were made to establish the parameter regime where long-range magnetic order gives way to a QSL. Whereas nearest-neighbor Heisenberg interactions on the triangular lattice support the $120^{\circ}$ magnetic order~\cite{Zhitomirsky2013}, a weak second-neighbor coupling is sufficient to suppress this order and drive the system toward a QSL state~\cite{Kaneko2014,Hu2015,Zhu2015,Bishop2015,Iqbal2016,Saadatmand2016,Bauer2017,Saadatmand2017}. The presence of multiple anisotropic interactions--a characteristic of the Yb$^{3+}$ compounds--lays out another route to the QSL~\cite{Zhu2018,Maksimov2018}. These two regimes, the second-neighbor isotropic exchange vs. nearest-neighbor anisotropic exchange, may in fact produce isomorphic QSL phases~\cite{Zhu2018}, but their exact nature and even the presence~\cite{Zhu2015,Bauer2017} or absence~\cite{Kaneko2014,Bishop2015,Iqbal2016} of a spin gap therein remain vividly debated. Experimental input is thus highly desirable, but requires a disorder-free material characterized down to mK temperatures, as magnetic interactions in the Yb$^{3+}$ oxides are of the order of 1\,K, and the ground-state regime is practically reached below 0.4\,K only~\cite{Li2015SR,Li2016PRL}.

\begin{figure}
\centering
\includegraphics[width=0.8\linewidth]{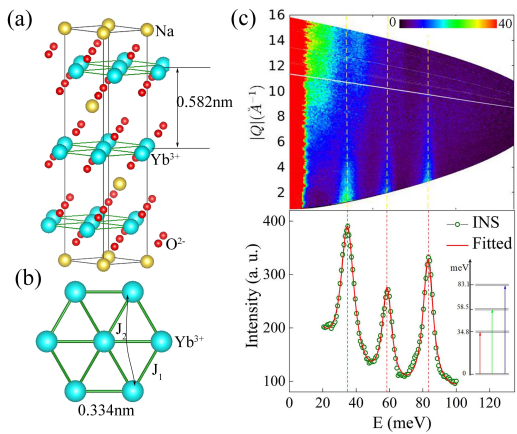}
\caption{(a) Stacking of the Yb$^{3+}$ triangular-layers along the $c$-axis. (b) Triangular layer formed by Yb$^{3+}$ cations. (c) Inelastic neutron scattering spectra $S(Q, \hbar\omega)$ at 5 K with $E_i$=150 meV. Energy dependence of the INS intensity at 5 K integrated in $Q$ over the range 3.9-4.4 \AA$^{-1}$. Inset depicts the CEF transitions from the ground-state Kramers doublet.}\label{fig:1}
\end{figure}

Na-based chalcogenides NaYbX$_2$ (X = O, S, Se) have recently come to the attention of researchers as disorder-free triangular antiferromagnets~\cite{Baenitz2018,Liu2018}. They feature layers of edge-sharing YbO$_6$ octahedra with the triangular arrangement of the magnetic Yb$^{3+}$ ions. These layers are separated by the well-ordered Na atoms, with the interlayer Yb--Yb distance of 5.82\,\r A, which is shorter than in YbMgGaO$_4$ (8.61\,\r A), but still significantly longer than the nearest-neighbor Yb--Yb distance of 3.34\,\r A within the triangular planes (see Fig.~\ref{fig:1}(a-b)).

Here, we confirm the absence of structural disorder and report persistent spin dynamics in NaYbO$_2$ down to at least 70\,mK. We also observe continuous excitations that are qualitatively similar to those predicted~\cite{Zhu2018} for the QSL state in triangular antiferromagnets. Our results set up NaYbO$_2$ as a new, disorder-free spin-liquid candidate, and shed light onto the physics of the spin-liquid state in triangular antiferromagnets. We demonstrate gapless nature of this state and the absence of simple power-law scaling for the specific heat. 

\textit{Absence of structural disorder.} Polycrystalline samples of NaYbO$_2$ were synthesized by a solid-state reaction as described in Ref.~\onlinecite{Hashinoto2003}. The absence of structural disorder was verified by synchrotron x-ray and neutron diffraction data~\cite{suppl} collected, respectively, at the ID22 beamline of the ESRF and at the WISH instrument~\cite{Chapon2011} at the ISIS facility. No signatures of site deficiency or disorder was observed in the structure refinements performed down to 1.5\,K. The high-resolution synchrotron data reveal very sharp peaks and exclude any extended defects that may occur in a layered compound~\cite{suppl}. At 10\,K, the atomic displacement parameter of Yb is below $10^{-3}$\,\r A$^2$ and excludes any off-center displacements that have been the most direct signature of structural randomness in YbMgGaO$_4$~\cite{Li2017PRL}.

In YbMgGaO$_4$, the structural disorder becomes most conspicuous in the CEF excitations that broaden and even show four peaks in the inelastic neutron spectrum~\cite{Li2017PRL}, instead of the three peaks expected for Yb$^{3+}$ with its $^2F_{7/2}$ multiplet split into four Kramers doublets by the trigonal crystal field~\cite{Gaudet2015,Ross2011}. The CEF excitations of NaYbO$_2$ were measured at 5\,K using the MERLIN spectrometer at ISIS operating with the incident energies of 90 and 150\,meV~\cite{Bewley2006}. As shown in Fig.~\ref{fig:1}(c), three sharp, resolution-limited CEF excitations are observed, as expected for Yb$^{3+}$. This ultimately proves the absence of structural disorder in our material. 

From the excitation energies and line intensities we extract the CEF parameters and the compositions of the four Kramers doublets~\cite{suppl}. It is worth noting that the excitation energies of 34.8, 58.5, and 83.1\,meV are not far from those reported for YbMgGaO$_4$ (39.4, 61.3, and 96.6\,meV, respectively~\cite{Li2017PRL}), reflecting similar local environments of Yb$^{3+}$ in both compounds. By contrast, NaYbS$_2$ bears all CEF excitations below 50\,meV~\cite{Baenitz2018}, likely in agreement with the less ionic nature of the Yb--S bonds. Similar CEF excitations in YbMgGaO$_4$ and NaYbO$_2$ indicate that on the level of single-ion physics the latter can be seen as a close analogue of the former, but with the structural disorder completely removed.

\begin{figure}[t]
\centering
\includegraphics[width=1\linewidth]{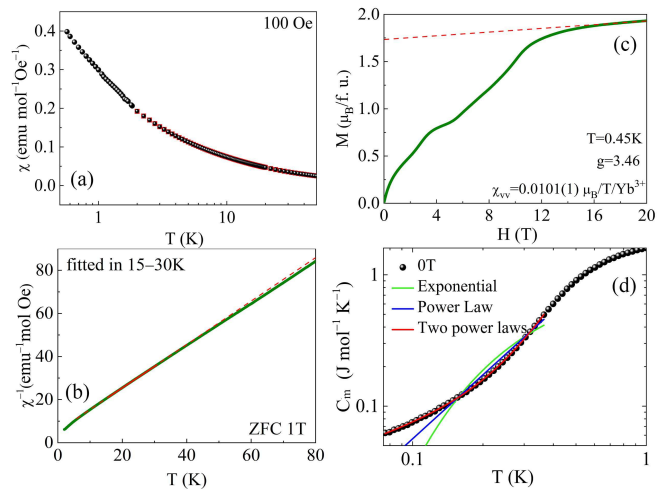}
\caption{(a) Temperature dependence of the magnetic susceptibility of NaYbO$_2$ down to 0.4 K. Zero-field-cooling and field-cooling curves are shown down to 1.8 K. (b) The Curie-Weiss fit to the inverse magnetic susceptibility in the range of 15-30 K after subtracting the Van Vleck term. (c) Isothermal magnetization curve measured at 0.45 K up to 60 T. The red line marks the Van Vleck contribution. (d) Zero-field specific heat of NaYbO$_2$ with the nuclear contribution subtracted~\cite{suppl}.}  \label{fig:2}
\end{figure}

\textit{Ground-state CEF doublet.} At low temperatures, the magnetic behavior of NaYbO$_2$ is fully determined by the ground-state Kramers doublet and can be described by an effective pseudospin-$\frac12$ Hamiltonian~\cite{Chen2016}. Indeed, inverse magnetic susceptibility measured using a SQUID magnetometer (Quantum Design, MPMS-7T) shows a change in slope around 70\,K, with the low-temperature linear part reflecting the Curie-Weiss behavior of pseudospins-$\frac12$ (see Fig.~\ref{fig:2}(a-b)). Nevertheless, the excited CEF levels produce a sizeable Van Vleck term $\chi_{vv}$. 

To determine the $\chi_{vv}$, we measured field-dependent magnetization at 0.45\,K using a triply compensated extraction magnetometer within a 65\,T short-pulse magnet at the National High Magnetic Field Laboratory, Los Alamos. As seen in Fig.~\ref{fig:2}(c), NaYbO$_2$ saturates around 16\,T. The magnetization increases linearly in higher fields, and its slope corresponds to the van Vleck term of $\chi_{vv}=0.0101$\,$\mu_B$/T$=0.00564$\,emu/mol/Oe~\cite{suppl}. After subtracting this Van Vleck contribution, we arrive at the saturation magnetization of 1.75\,$\mu_B$/f.u. Magnetic susceptibility with $\chi_{vv}$ subtracted follows the Curie-Weiss law with the effective moment of $\mu_{\rm eff}=2.84(2)$\,$\mu_B$. Both values are comparable favorably to 1.5\,$\mu_B$/f.u. and 2.60\,$\mu_B$ expected from the powder-averaged $\bar g=3.00$ calculated based on the CEF fit.

\textit{Low-temperature thermodynamics.} The susceptibility fit also yields the Curie-Weiss temperature $\Theta=-5.64(1)$\,K. Its negative value confirms antiferromagnetic interactions between the Yb$^{3+}$ pseudospins. The absolute value is about two times larger than in YbMgGaO$_4$ ($\Theta_{\|}=-1.5$\,K, $\Theta_{\perp}=-2.7$\,K~\cite{Li2015PRL}) and corroborates the two-fold increase in the saturation field from 8\,T in YbMgGaO$_4$ to 16\,T in NaYbO$_2$. The simple estimate $\Theta=-3\bar J_1/2$ for a triangular antiferromagnet yields $\bar J_1\simeq 3.8$\,K, the energy scale of nearest-neighbor exchange couplings in NaYbO$_2$. 

To probe low-temperature thermodynamics, we measured specific heat ($C_p$) of NaYbO$_2$ using the PPMS (Quantum Design) down to 0.4\,K and a home-built dilution-fridge setup down to 70\,mK~\cite{suppl}. Magnetic contribution to the $C_p$ becomes visible below 8\,K and shows a broad maximum around 1\,K~\cite{suppl}. At even lower temperatures, magnetic contribution decreases without showing any signatures of a magnetic transition (see Fig.~\ref{fig:2}(d)). This suggests the absence of long-range magnetic order and the possibility of a spin-liquid state confirmed by the muon spin rotation ($\mu$SR) measurement below. 

By taking NaLuO$_2$ as the isostructural non-magnetic compound, we confirmed that lattice contribution to the $C_p$ is negligible below 2\,K, whereas the nuclear contribution was subtracted by systematic measurements in weak applied fields~\cite{suppl}. The remaining, magnetic contribution $C_m(T)$ will usually take the exponential form, $e^{-\Delta/T}$, or the form of a power law, $T^p$, for gapped and gapless excitations, respectively. However, neither term accounts for our experimental data (Fig.~\ref{fig:2}(d)). We tentatively fit $C_m(T)$ with two power laws, $aT^p+bT^q$, where $p\simeq 2.9$ resembles the $T^3$ contribution of magnons in a long-range-ordered antiferromagnet, and $q\simeq 0.5$ reflects a sublinear behavior, which is remotely similar to the power-law behavior in YbMgGaO$_4$~\cite{Li2015SR}. At first glance, a combination of the two different contributions could be seen as an effect of sample inhomogeneity, but a weak applied field restores the simpler $T^{2.2}$ power-law behavior~\cite{suppl}, suggesting an intrinsic nature of the two power laws observed in zero field. We conclude that the low-energy excitations in NaYbO$_2$ are neither gapped nor magnon-like. They are also distinct from the low-energy excitations in YbMgGaO$_4$ that showed~\cite{Li2015SR} the robust $T^{0.7}$ power law characteristic of the "spinon metal" of a U(1) quantum spin liquid~\cite{Motrunich2005}.

Before going further, we note in passing that the magnetization curve of NaYbO$_2$ measured at 0.45\,K shows a plateau between 4 and 5\,T at about one half of the saturation magnetization as shown in Fig.~\ref{fig:2}(c). Such a $\frac12$-plateau contrasts with the $\frac13$-plateau typically observed in Heisenberg and XXZ triangular antiferromagnets~\cite{Starykh2015} and may be indicative of a more complex interaction regime. Moreover, field-induced phase transitions should occur in NaYbO$_2$, but they go beyond the scope of our present manuscript and will be addressed in future studies.

\begin{figure}
\centering
\includegraphics[width=0.9\linewidth]{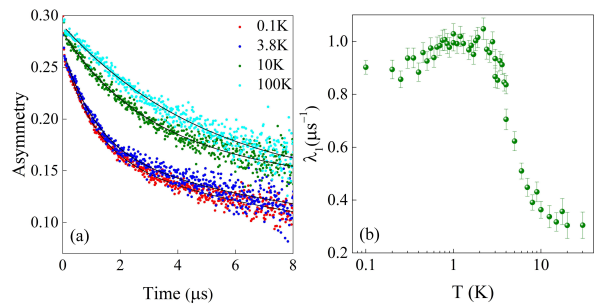}
\caption{ (a) ZF $\mu$SR spectra collected at various temperatures. The solid lines represent the fit using function (1). (b) Temperature dependence of the muon spin relaxation rate for the ZF $\mu$SR spectra.}\label{fig:3}
\end{figure} 

\textit{$\mu$SR data.} Our heat-capacity data exclude long-range magnetic ordering above 70\,mK, yet spin freezing would have no immediate effect on the specific heat. A direct probe of spin dynamics is thus essential to identify the QSL. To this end, we carried out zero-field (ZF) and longitudinal-field (LF) $\mu$SR measurement down to 100\,mK at the MuSR spectrometer (ISIS). 

Four representative ZF $\mu$SR spectra are shown in Fig.~\ref{fig:3}(a). They reveal neither oscillations nor a drastic drop in the initial asymmetry, indicative of the long-range magnetic order, but instead show signatures of persistent spin dynamics, in particular the lack of polarization recovery to $\frac13$ of the initial value rules out the presence of static random fields. We further analyze these data by fitting the muon spectra with two exponential components:
\begin{equation}
A(t)=A_0[f_1exp(-\lambda_1t)+(1-f_1)exp(-\lambda_2t)]+B_0
\end{equation} 
where A$_0$ and B$_0$ denote the initial asymmetry and the constant background, respectively, $\lambda_1$ and $\lambda_2$ represent the muon spin relaxation rates for muons implanted at two sites near O$^{2-}$, f$_1$ stands for the fraction of the first component.  The fitted ZF $\mu$SR relaxation rate $\lambda_1$, $\lambda_2$ and f$_1$ as a function of temperature are shown in \cite{suppl}. f$_1$ shows temperature-independent behavior with its value close to 0.5, suggesting the same population at the two muon sites. Below we will discuss the electronic dynamics in terms of $\lambda_1$ since it is significantly larger than $\lambda_2$.

Temperature dependence of $\lambda_1$ tracks the onset of correlations between the Yb$^{3+}$ pseudospins. As seen in Fig.~\ref{fig:3}(b). Above 10\,K, NaYbO$_2$ is paramagnetic with a smaller and temperature-independent $\lambda_1$. The increase in $\lambda_1$ below 10\,K is accompanied by the growing magnetic contribution to the specific heat, whereas the second temperature-independent regime below 2\,K indicates the onset of the spin-liquid state. To prove the dynamic nature of the relaxation in this temperature range, we performed the LF experiment at 1.5\,K. Should the relaxation arise from a weak static field, the size of this field is $B_{\rm loc}=\lambda/\gamma_{\mu}\simeq 1.2$\,mT, where $\gamma_{\mu}=135.5\times 2\pi$\,s$^{-1}$\,$\mu$T$^{-1}$ is the gyromagnetic ratio for muons. Our LF data show that the relaxation persists in much higher fields, thus proving the dynamic nature of the Yb$^{3+}$ pseudospins~\cite{suppl}. 

We also note that the characteristic evolution of $\lambda_1$, its increase below 10\,K and the saturation below 2\,K, takes place at about twice higher temperatures compared to YbMgGaO$_4$~\cite{Li2016PRL}. This further supports our conclusion on the twice stronger exchange couplings in NaYbO$_2$. 

\begin{figure}
\centering
\includegraphics[width=1\linewidth]{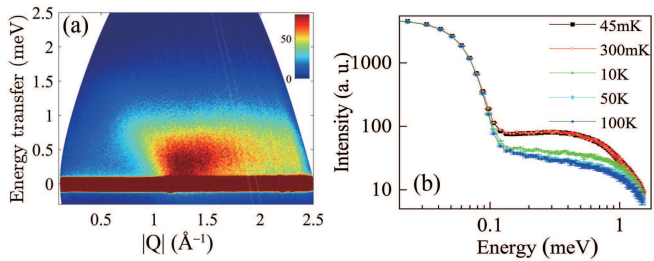}
\caption{(a) INS spectra measured at 45 mK with the incident energies $E_i$=3.7 meV. (b) Energy dependence of the integrated cut along $Q$ in the range of 1.2-1.5 \AA$^{-1}$.}\label{fig:4}
\end{figure}

\textit{Low-energy excitations.} The most interesting property of a spin liquid is arguably its excitation spectrum. We probed the low-energy excitations of NaYbO$_2$ at the cold-neutron multi-chopper LET spectrometer (ISIS)~\cite{Bewley2011} at 45\,mK using incident neutron energies of 1.46, 3.7, and 7.52\,meV. Spectral weight observed above 0.8\,\r A$^{-1}$ is continuously distributed in both energy ($E$) and momentum ($Q$) and extends to about 1\,meV (see Fig.~\ref{fig:4}(a)). This low-energy spectral weight concentrates around $Q\simeq 1.25$\,\r A$^{-1}$ corresponding to the $K$-point of the Brillouin zone with the reciprocal-lattice vector ($\frac13$, $\frac13$, 0). A $120^{\circ}$ magnetic order will lead to a Bragg peak at the same position, but the excitation spectrum is clearly different from the spin-wave spectrum of such an ordered state~\cite{suppl}. The energy dependence over the constant-$Q$ cut suggests that the excitations reach low energies down to the elastic line that becomes prominent below 0.1\,meV. This further confirms gapless nature of the magnetic excitations in NaYbO$_2$ (see Fig.~\ref{fig:4}(b)). 

The spectrum of NaYbO$_2$ bears close similarities to that of YbMgGaO$_4$, where no spectral weight was observed at low $Q$, around the zone center. All the spectral weight is concentrated in the vicinity of the zone boundary, in agreement with our observation of the spectral weight above 0.8\,\r A$^{-1}$ only. A further, and more subtle feature of YbMgGaO$_4$, is the re-distribution of the spectral weight upon heating that signals the formation of distinct excitations at energies above and below $\bar J_1$~\cite{Li2017NC}. No such effect is seen in NaYbO$_2$. Moreover, magnetic excitations in YbMgGaO$_4$ extend to the much higher energy of 2\,meV, despite the fact that $\bar J_1$ is twice smaller than in NaYbO$_2$.

\textit{Discussion.} NaYbO$_2$ shows strong similarities to the widely studied triangular spin-liquid candidate YbMgGaO$_4$ except for the absence of structural disorder. Both materials entail the trigonally distorted YbO$_6$ octahedra. The CEF excitations of Yb$^{3+}$ occur at about the same energies, and the compositions of the ground-state Kramers doublets are similar too. Exchange couplings differ by a factor of 2, though. This change is accompanied by a reduction in the Yb--O--Yb bridging angle from $99.4^{\circ}$ in YbMgGaO$_4$ to $95.7^{\circ}$ in NaYbO$_2$. 

The similarity between the two materials gives us an interesting opportunity to explore which of the effects reported for YbMgGaO$_4$ appear in the disorder-free case. Gapless ground state is retained in NaYbO$_2$. On the other hand, neither the simple power-law behavior of the low-temperature specific heat~\cite{Li2015SR}, nor the energy separation of the spin excitations~\cite{Li2017NC} have been observed. Moreover, the spin excitations of NaYbO$_2$ extend to much lower energies despite the twice larger $\bar J_1$. This goes in line with the theory of Ref.~\onlinecite{Kimchi2018} that explains both effects in terms of quenched disorder in a valence-bond solid. These effects are probably not generic to the spin-liquid state of triangular antiferromagnets.

Rau and Gingras~\cite{Rau2018} computed anisotropic nearest-neighbor exchange couplings for several triangular geometries inspired by the possible local structures of YbMgGaO$_4$. By extrapolating their results to the Yb--O--Yb angle of $95.7^{\circ}$ in NaYbO$_2$, we may expect that at least the nearest-neighbor coupling $\bar J_1$ in this material is very close to the Heisenberg limit. The second-neighbor interaction $\bar J_2$ should be then operative in order to stabilize a QSL. 
%which is inconsistent with the spinon Fermi surface QSL state. 
Such a scenario may be envisaged if, for example, $\bar J_2$ is less sensitive to the structural geometry than $\bar J_1$. With $\bar J_2/\bar J_1=0.18(7)$ reported for YbMgGaO$_4$~\cite{Zhang2018}, the increase in $\bar J_1$ accompanied by a small change in $\bar J_2$ will drive the system directly into the QSL phase~\footnote{One could argue that $\bar J_2$ increases as well, but $\bar J_1\simeq 3.8$\,K will then lead to $\bar J_2$ of nearly 1\,K, which is hard to reconcile with the second-neighbor Yb--Yb distance of 5.82\,\r A and the strongly localized nature of $f$-electrons in Yb$^{3+}$.}. Another interesting observation is that static structure factor calculated for this QSL phase peaks at the $K$-point of the Brillouin zone~\cite{Zhu2018} in agreement with the accumulation of the low-energy spectral weight at $Q\simeq 1.25$\,\r A$^{-1}$ observed in our experiment (see Fig.~\ref{fig:4}), whereas YbMgGaO$_4$ shows a larger low-energy spectral weight at the $M$-point with the reciprocal-lattice vector ($\frac12$, 0, 0)~\cite{Paddison2017}. All these arguments give a strong envision that the spin-liquid state observed in NaYbO$_2$ is the QSL phase of triangular antiferromagnets. Its gapless nature and unusual sensitivity to the magnetic field open prospects for future studies theoretically and experimentally. 

\textit{Conclusions.} The disorder-free NaYbO$_2$ gives the most direct experimental access to the spin-liquid physics of triangular antiferromagnets. Thermodynamic measurements and muon spectroscopy indicate the absence of magnetic order and persistent spin dynamics down to at least 70\,mK. An excitation continuum is observed, with the spectral weight accumulating around the $K$-point, as expected in the QSL phase(s) driven by the exchange anisotropy or second-neighbor coupling on the triangular lattice~\cite{Zhu2018}. The spin-liquid state of NaYbO$_2$ is gapless with a non-trivial low-temperature evolution of the specific heat, which does not follow the spinon scenario originally proposed for YbMgGaO$_4$. 

\acknowledgements
LD acknowledges support from the Rutherford International Fellowship Programme (RIFP). This project has received funding from the European Union's Horizon 2020 research and innovation programme under the Marie Sk\l{}odowska-Curie grant agreements No.665593 awarded to the Science and Technology Facilities Council. RDJ acknowledges financial support from the Royal Society. The work in Augsburg was supported by the German Science Foundation under TRR80. LD thanks G. Stenning for his help during the thermodynamic measurements in the Materials Characterisation Laboratory and P. Biswas during the muon data collection at the ISIS facility. AT thanks Yuesheng Li, Mayukh Majumder, Michael Baenitz, and Liviu Hozoi for useful discussions, and ESRF for providing the beamtime at ID22.

%\begin{thebibliography}{}%

\end{document}